\documentclass[10pt,letterpaper]{article}
\usepackage{opex3}
\usepackage{amssymb}

\begin{document}

\title{Frequency Stabilization of a 369 nm Diode Laser by Nonlinear Spectroscopy of Ytterbium Ions in a Discharge}

\author{Michael W Lee,$^{{1}}$ Marie Claire Jarratt,$^{{1}}$ Christian Marciniak,$^{{1}}$ and Michael J Biercuk$^{{1,*}}$}

\address{$^1$ARC Centre for Engineered Quantum Systems, School of Physics, The University of Sydney, NSW 2006, Australia
National Measurement Institute, Lindfield NSW 2070, Australia
}

\email{*biercuk@physics.usyd.edu.au} 



\begin{abstract}
We demonstrate stabilisation of an ultraviolet diode laser via Doppler free spectroscopy of Ytterbium ions in a discharge.  Our technique employs polarization spectroscopy, which produces a natural dispersive lineshape whose zero-crossing is largely immune to environmental drifts, making this signal an ideal absolute frequency reference for Yb$^+$ ion trapping experiments.  We stabilise an external-cavity diode laser near $369$ nm for cooling Yb$^+$ ions, using amplitude-modulated polarisation spectroscopy and a commercial PID feedback system.  We achieve stable, low-drift locking with a standard deviation of measured laser frequency $\sim400$ kHz over $10$ minutes, limited by the instantaneous linewidth of the diode laser. These results and the simplicity of our optical setup makes our approach attractive for stabilization of laser sources in atomic-physics applications.
\end{abstract}



\section{Introduction}

Experiments in atomic and quantum physics with trapped ions require very specific frequencies of laser radiation for ion cooling, state preparation and readout, and enacting quantum gates \cite{metcalf1999lca}. For commonly used species such as Ytterbium, frequency stability of the lasers is required within at most a few MHz of the desired value, set by the natural linewidth of the relevant atomic transition.  This requirement usually necessitates active feedback based on a stable frequency reference due to drifts and environmental perturbations such as pressure and temperature fluctuation.  These frequency references can be broadly categorized into \emph{absolute} frequency references, such as an atomic or molecular transition \cite{arie1992afs,wallard1972fso,corwin1998fsd,ratnapala2004lfl} or calibrated \emph{relative} frequency references, such as an optical cavity \cite{hansch1980lfs,drever1983lpa} or interferometers in a wavelength meter. The particular frequencies in question depend on the ion species used and are often in the ultraviolet (UV) part of the spectrum.

Advances in laser technology have simplified the process of generating the relevant frequencies but have led to new complications in laser stabilization, for example due to directly generating light in frequency ranges where few absolute frequency references exist.  For instance, the recent development of UV laser diodes capable of operating at $369.5$ nm in an external-cavity diode laser (ECDL) allow for the direct generation of this wavelength, which corresponds to the $^2S_{1/2}\rightarrow^2P_{1/2}$ cooling transition in Ytterbium ions \cite{kielpinski2006lco,nguyen2010ett}. These ECDLs are attractive light sources as they offer a broad wavelength tuning range in a compact package with low cost and technical complexity compared to previous sources.  

In past approaches, the required wavelengths were predominantly generated by nonlinear frequency conversion of a visible or near infrared laser source \cite{wunderlich2003qma,gill1995mot,sugiyama1995lco}. This allowed feedback stabilization on the long-wavelength sources, via e.g. spectroscopy of Iodine vapor which has many molecular transitions in the visible and near infrared \cite{arie1993isa,gerstenkorn1985dot}. However, this broad range of absolute frequency references is more limited in the UV. 

In this work we demonstrate a laser stabilization technique directly on a UV source that is based on performing H\"{a}nsch-Coulliaud nonlinear polarization spectroscopy \cite{ratnapala2004lfl,hansch1980lfs,demtroder2008ls,wieman1976dfl} on Ytterbium ions in the discharge generated in a hollow cathode lamp (HCL). Our work provides a Doppler free absolute frequency reference for a large number of relevant Ytterbium ion isotopes near $369.5$ nm, and builds on previous experiments deriving dispersive error signals via  dichroic atomic vapor laser locking (DAVLL) \cite{corwin1998fsd}, optogalvanic spectroscopy \cite{nguyen2010ett,streed2008fso,petrasiunas2012oso} and modulation transfer spectroscopy \cite{wang2009fso} in the same hollow cathode lamp. Compared with previous techniques, however, our approach provides enhanced robustness against ambient environmental fluctuations, and does not rely on the stability of an external magnetic field or power supply in order to produce a dispersive signal whose zero-crossing is stable in frequency. 

The remainder of this manuscript is organized as follows. In Section \ref{sec:pol_spec} we outline the basic physical phenomenon employed in generating the nonlinear spectroscopic signal. This is followed in Section \ref{sec:expt_method} by a detailed description of our optical system, and demonstration of both nonlinear spectroscopy and laser-frequency stabilization to the dispersive signal in Sections \ref{sec:spect_meas} and \ref{sec:laser_lock} respectively.

\section{Polarization spectroscopy of Ytterbium ions}\label{sec:pol_spec}

Polarization spectroscopy is a Doppler-free technique based on counter-propagating pump and probe beams and is closely related to saturated absorption spectroscopy \cite{demtroder2008ls}.  A spectroscopic signal with a natural dispersive lineshape and sub-Doppler resolution is derived by examining the circular dichroism induced by passage of the pump laser through a spectroscopic medium. This approach brings the additional benefit of immunity to common mode laser amplitude fluctuations as the dispersive signal is derived from the orthogonal polarization components of a single probe beam.  The principle of polarization spectroscopy will be briefly reviewed here following \cite{demtroder2008ls,pearman2002pso}. 

An induced circular dichroism in the medium corresponds to a difference in the absorption coefficients experienced by the left and right hand circular polarization components of the probe wave $\Delta \alpha = \alpha^{+} - \alpha^{-} \neq 0$. By the Kramers-Kronig dispersion relation, this corresponds to a difference in refractive index given by \cite{demtroder2008ls}:
\begin{eqnarray}\label{eq:KK}
n^{+} - n^{-} = \Delta n = \frac{c}{\omega_{0}}\frac{\Delta \alpha_{0}x}{1+x^{2}}
\end{eqnarray}
where $x = (\omega_{0}-\omega)/(\Gamma/2)$ is the detuning from the transition frequency in units of half the transition line width and $\Delta \alpha_{0}$ is the difference in absorption coefficients at the resonant frequency.
\\
\\
The probe wave travelling along the $\hat{z}$-axis and linearly polarized at an angle $\phi$ with respect to the $\hat{x}$-axis can be written as $E = E^{+} + E^{-}$, where:
\begin{eqnarray}
E^{+} =& E^{+}_{0}e^{i(\omega t - k^{+}z)},\quad E^{+}_{0} = \frac{1}{2}E_{0}e^{-i\phi}(\hat{x}+i\hat{y})\\
E^{-} =& E^{-}_{0}e^{i(\omega t - k^{-}z)},\quad E^{-}_{0} = \frac{1}{2}E_{0}e^{+i\phi}(\hat{x}-i\hat{y})
\end{eqnarray}
After passing through the spectroscopic medium (here the HCL) with length $L$ the phase difference between the two polarization components is given by:
\begin{eqnarray}
\Delta \phi = (k^{+}+k^{-})L = \frac{\omega L \Delta n}{c}
\end{eqnarray}
and the amplitude difference by:
\begin{eqnarray}
\Delta E = \frac{E_{0}}{2} \left[ e^{-(\alpha^{+}/2)L} - e^{-(\alpha^{-}/2)L}\right]
\end{eqnarray}
The probe wave after the HCL is then:
\begin{eqnarray}
E^{+} =& E^{+}_{0} e^{i[\omega t - k^{+}L + i(\alpha^{+}/2)L]}\\
E^{-} =& E^{-}_{0} e^{i[\omega t - k^{-}L + i(\alpha^{-}/2)L]}
\end{eqnarray}
giving the total electric field of the probe wave as:
\begin{eqnarray}
E = \frac{E_{0}}{2} e^{i\omega t} e^{-i[ \frac{\omega L n}{c} - \frac{i\alpha L}{2} ]} \left[e^{-i\phi} e^{-i\Omega}(\hat{x}+i\hat{y})  + e^{+i\phi}e^{+i\Omega}(\hat{x}-i\hat{y}) \right]z
\end{eqnarray}
where $n = \frac{1}{2}(n^{+}-n^{-})$, $\alpha = \frac{1}{2}(\alpha^{+}-\alpha^{-})$ and
\begin{eqnarray}
\Omega = \frac{\omega L \Delta n}{2c} - i\frac{L\Delta \alpha}{4}
\end{eqnarray}
This derivation ignores any additional contribution to an imbalance in the circular polarisation due to experimental hardware, here for instance, birefringence in the windows of the HCL, which is assumed to be negligible.

We thus see that there is a shift in the relative phases and amplitudes of the orthogonal circular polarisations of light in the probe beam, providing a means to determine when the laser frequency is near a spectroscopic resonance.  The polarisation spectroscopy measurement is made by projecting into the linear polarisation basis and taking the difference of the horizontal and vertical polarization components of the probe. Physically the probe beam is split into these two components using bulk polarization optics, and is incident upon a balanced pair of photodetectors with output signal proportional to the difference in intensity between the horizontal and vertical components:
\begin{eqnarray}
I = I_{V} - I_{H} = I_{0}e^{-\alpha L}\cos\left(2\phi + \frac{\omega L \Delta n}{c} \right)
\end{eqnarray}
This equation can be re written by substituting in the Kramers-Konig relation (Eq. \ref{eq:KK}). Furthermore, the signal will be maximised when $\phi = \frac{\pi}{4}$, corresponding to the I$_2$ Stokes parameter of the probe beam \cite{ratnapala2004lfl}. This allows the simplification $\cos(\pi/ 2 + \theta)\to-\sin(\theta)$, and the small angle approximation $-\sin(\theta) \approx -\theta$. The result is:
\begin{eqnarray}\label{eq:Intensity}
I = -I_{0}e^{-\alpha L}\Delta \alpha_{0}L \frac{x}{1 + x^{2}}
\end{eqnarray}

The form of this equation yields a dispersive type curve as a function of detuning from an absorption peak; i.e. it is an anti-symmetric function with a zero crossing at the center of the absorption peak. This functional form provides an excellent error signal for laser locking to an absorption feature.

\begin{figure}[htbp]
\centering\includegraphics[width=12cm]{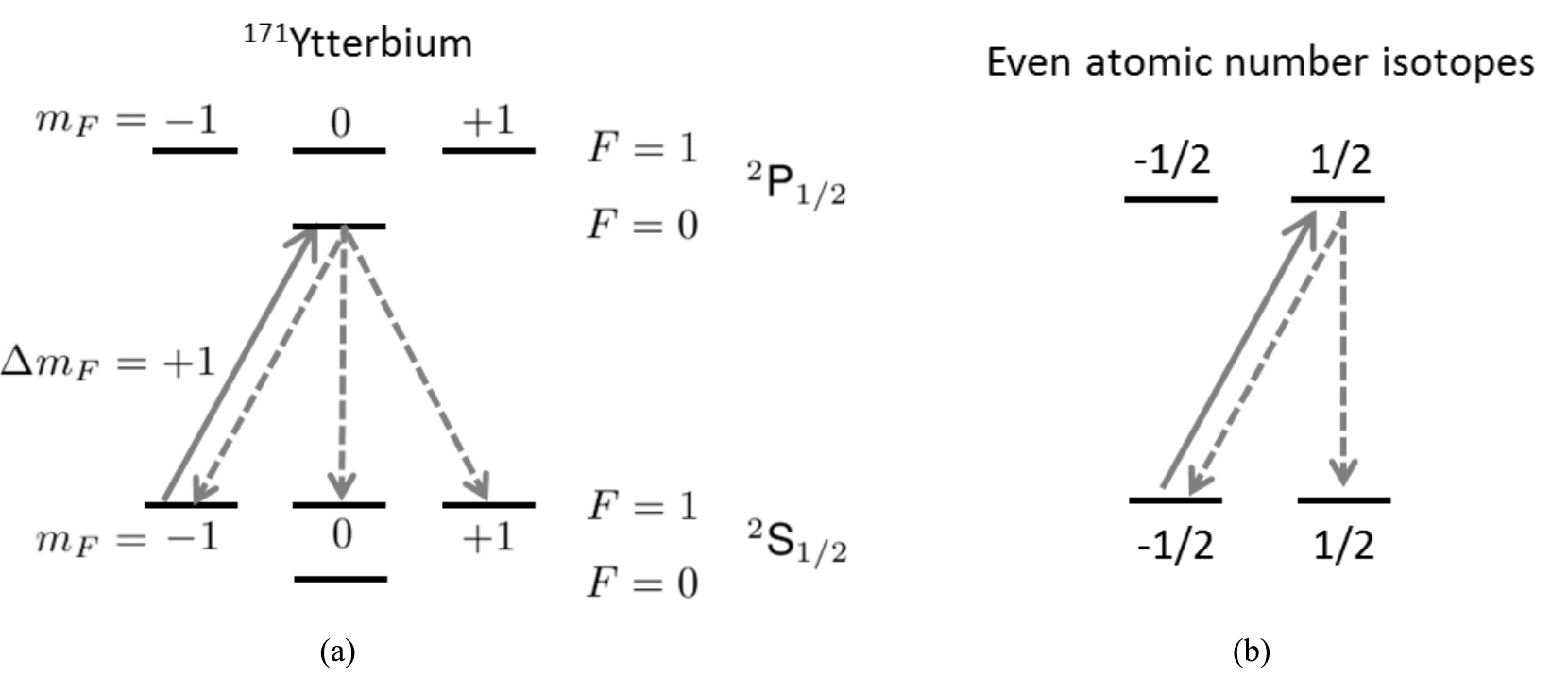} 
\caption{The structure of the $^2S_{1/2} \rightarrow ^2P_{1/2}$ transition for (a): the $^{171}$Yb$^+$ ion, showing the hyperfine structure and the optical pumping scheme for the $F=1\rightarrow F'=0$ transition with $\sigma^+$ light on resonance. With pure $\sigma^+$ light the only allowed transition is the $F=1, m_F=-1 \rightarrow F'=0, m_{F'}=0$ transition, while decay from the upper level can be into any of the three lower sublevels. This results in depletion of the population in the $F=1, m_F=-1$, saturating the absorption for just the $\sigma^+$ light and giving rise to the circular dichroism required for polarization spectroscopy. (b): The even atomic number isotopes do not have hyperfine structure, instead the circular dichroism results from optical pumping of the Zeeman states (labeled by -1/2 and 1/2) of the upper and lower levels. In a similar way this results in saturation of the absorption only for $\sigma^+$ light.}
\label{fig:yb_lvl_struct}
\end{figure}

One of the key parameters in Eq. \ref{eq:Intensity} is $\Delta\alpha _0$, the differential line center absorption between the left and right circular components of the polarization. For Yb$^+$ ions in the HCL discharge this occurs due to optical pumping of the $m_F$ sublevels of the $^2S_{1/2}$, $F=1$ level of the $^{171}$Yb$^+$ ion. This process is depicted in Fig. \ref{fig:yb_lvl_struct}(a), which shows the structure of the hyperfine levels involved in the $^2S_{1/2} \rightarrow ^2P_{1/2}$ transition. The circularly polarized pump beam used to induce the circular dichroism in the ions corresponds to $\sigma^+$ or $\sigma^-$ light depending upon the orientation of the ambient magnetic field and any magnetization of the hollow cathode material.

Here we are interested in the $F=1\rightarrow F'=0$ transition, assuming the circularly polarized pump beam corresponds to $\sigma^+$ light at the transition frequency.  We are then restricted to the excitation $m_F = -1 \rightarrow m_F' = 0$ due to the requirement that $\Delta m_F = +1$ for $\sigma^+$ light. An ion in the excited state decays obeying the magnetic quantum number selection rule $\Delta m_F=0,\pm1$ and thus can decay into any of the three $m_F$ levels of the ground state. This results in a depleted population in the $m_F = -1$ sublevel and an increased population in the other sublevels because of the asymmetry between the excitation and decay pathways. This unequal population distribution, or optical pumping, gives the difference in absorption, $\Delta\alpha_0$, between the $\sigma^+$ and $\sigma^-$ components of any incident light because the absorption is saturated for the $\sigma^+$ component but not for the $\sigma^-$ component. It should be noted that there is also a decay path from the $^2P_{1/2}$ level to the metastable $^2D_{3/2}$ level, where the ion is lost from our measurement, however this is not expected to have a significant detrimental effect due to collisional relaxation and the limited lifetime of individual ions in the discharge plasma.

Even-atomic-number isotopes of Yb$^+$ have zero nuclear magnetic moment and hence no hyperfine structure. In this case we simply consider the $-1/2$ and $+1/2$ Zeeman levels in both the ground and excited states with the $\sigma^+$ polarized light driving transitions from the $-1/2$ level of the ground state to the $+1/2$ level of the excited state. This is depicted schematically in Fig. \ref{fig:yb_lvl_struct}(b). Spontaneous emission from the excited state drops the electron into either of the two ground state levels resulting in optical pumping of the $+1/2$ state of the ground level.
 
In either of the above cases the process of optical pumping in the ion discharge of the HCL via the pump beam provides the circular dichroism required to generate the dispersive signal.  While the Zeeman sublevels may be split by the ambient magnetic field, we note that the generation of the dispersive lineshape, and in particular its zero-crossing, is not sensitively dependent on the magnetic field strength. The central transition frequency seen by the probe beam will be the average frequency of the transitions between $\left|F=1,m_F=-1\right\rangle\rightarrow \left|F'=0,m_F=0\right\rangle$ and $\left|F=1,m_F=1\right\rangle\rightarrow \left|F'=0,m_F=0\right\rangle$, so for small magnetic fields where the splitting of the stretched states is linear and symmetric and the $m_{F}=0$ Zeeman sublevel is not shifted, the zero crossing of the dispersive polarisation spectroscopy signal is minimally affected by ambient magnetic fields.

\section{Experimental Method}\label{sec:expt_method}

\begin{figure}[htbp]
\centering\includegraphics[width=8.5cm]{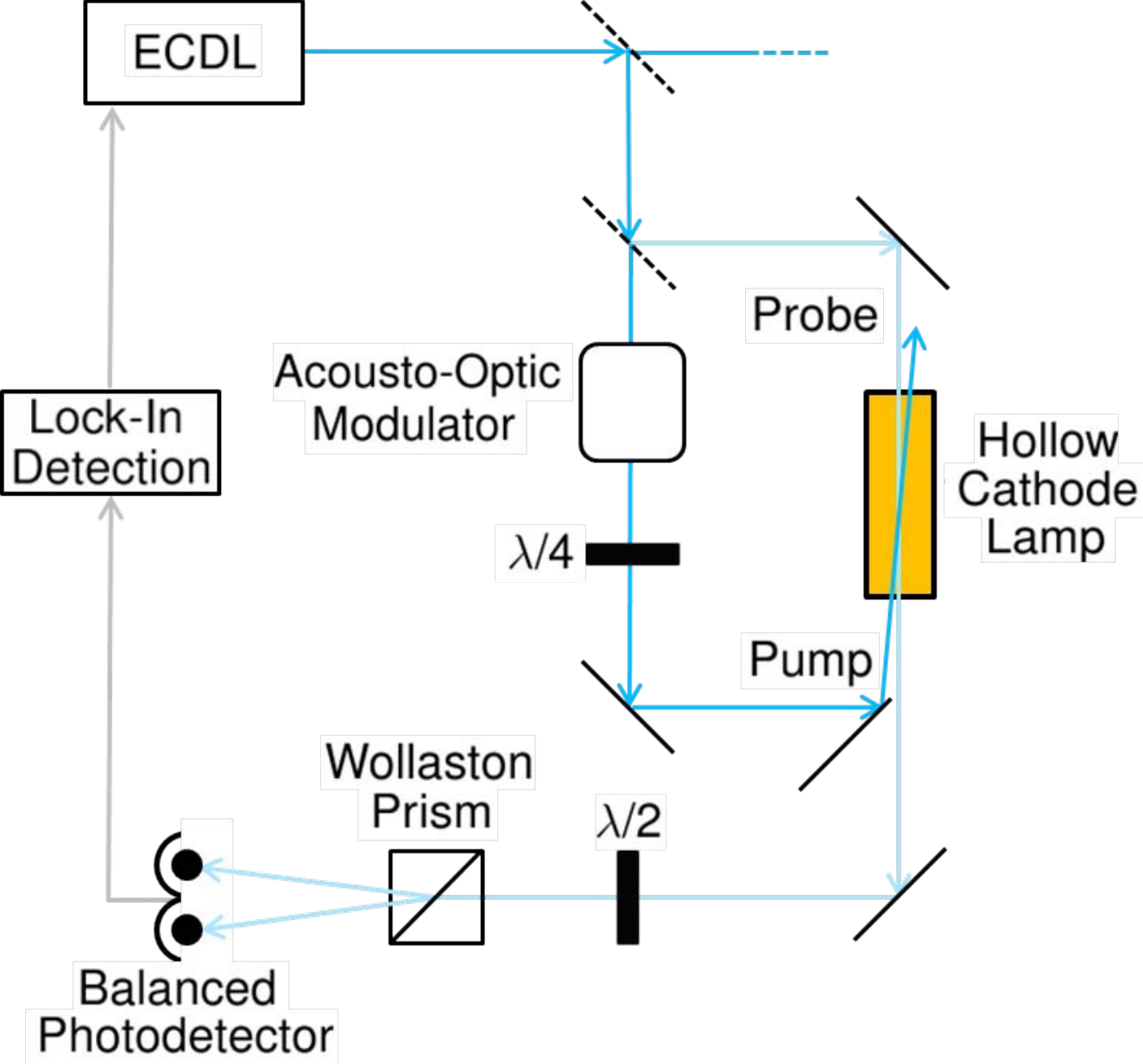}
\caption{A schematic of the setup used for the polarization spectroscopy measurements and laser locking. Light from the ECDL is split into a main beam for use in the ion trap and a beam for the laser locking system ($<$1 mW). The laser locking beam is then split into pump and probe beams, each initially with vertically oriented linear polarization. The pump beam is passed through an AOM, which is used to apply a square wave amplitude modulation for the purpose of lock-in detection, then the first order diffracted beam is passed through a quarter waveplate to change the polarization from linear to circular before being directed through the Yb$^+$ discharge in the HCL. The probe beam is directed through the HCL at a small angle to the pump beam to allow the beams to be separated, the beams are aligned so that they overlap in the Yb$^+$ and lenses are used to focus both beams in the center of the HCL and increase the interaction. After passing through the HCL the probe beam is passed through a half waveplate to rotate the vertical linear polarization to 45$^\circ$, a polarization beamsplitter (Wollaston prism) then separates the horizontal and vertical components and these components are then incident on a balanced pair photodetector. The difference signal from the balanced pair photodetector contains the polarization spectroscopy signal, however lock-in detection is used to improve the signal to noise ratio. This signal can be used as the error signal for frequency locking the ECDL using the internal servo in the MOGLabs DLC001 diode laser controller.}
\label{fig:expt_schematic}
\end{figure}

A schematic diagram of our experimental setup is shown in Fig. \ref{fig:expt_schematic}; the key components of the setup are the $369.5$ nm external-cavity diode laser (ECDL) and the Ytterbium HCL. The ECDL was a custom-modified commercial MOGLabs ECD001 with a Nichia NDU1113E $20$ mW diode selected for short-wavelength operation and mounted in the Littrow configuration \cite{ricci1995acg}. The laser has a maximum output power of $\sim6$ mW, operates near $369.5$ nm at room temperature, and allows a mode-hop-free tuning range of $\sim5$ GHz with a piezo element. The HCL is a Hamamatsu L2783 series see-through hollow cathode lamp with a Ytterbium cathode and Ne buffer gas. The rated lifetime of the lamp is $>5000$ mAh (reports indicate a useful life of $\sim30000$ mAh \cite{streed2008fso}).  Typical operating current is $10$ mA, supplied by a low-noise Kepco BHK-1000-40MG power supply, operating at approximately $100-200$ V.

In our experimental setup a linearly polarized beam from the ECDL is picked off for the polarization spectroscopy measurements and laser frequency stabilization while the remainder of the optical power is sent towards our main experimental apparatus.  The interrogating light is passed through a plate beamsplitter to create unbalanced pump and probe beams.  The pump is sent through an acousto-optic modulator (AOM) with the first diffracted order passed through a quarter wave plate and subsequently the HCL. Neutral density filters could be added in this beamline to adjust the pump power. 

The probe beam is directed through the HCL counterpropagating relative to the pump in order to narrow the velocity classes simultaneously interacting with both beams, hence permitting sub-Doppler feature resolution. The beams overlap in the centre of the HCL but cross at a shallow angle so that they may be physically separated.  Care is taken to prevent the beam waist from overlapping the electrode, as photoemission of electrons from the electrodes is a documented source of noise in such spectroscopic measurements \cite{kumar1995poe,zhechev2005ict}. After passing through the HCL the probe beam is sent through a half wave plate angled such that it would nominally (in the absence of the ion discharge) rotate the polarization from an initial vertical linear polarization to $45^\circ$ linear polarization. This beam is then passed through a Wollaston prism, which spatially separates the horizontal and vertical linearly polarized components of the beam. 

Circular dichroism induced by the interaction of the pump beam with the ion plasma in the HCL produces a net imbalance between the amplitude of the horizontally and vertically polarized components of the probe beam which is detected using a balanced photodetector.  This configuration corresponds to measuring the I$_2$ Stokes parameter \cite{ratnapala2004lfl} and ensures a zero background signal away from an atomic resonance. The required probe-beam power is largely determined by the quantum efficiency of the photodetectors in use. In our experiments we used a MOGLabs PDD001-$\pm10^\circ$-400-1100nm balanced pair photodetector with specified wavelength range of 400 nm - 1100 nm, hence at 369.5 nm we required a large probe power of 510 $\mu$W for our measurements; this could be much lower with a UV optimized detector.

In our measurement system we incorporate the AOM in order to modulate the pump beam for the purpose of lock-in detection of our dispersive signal. Although not strictly necessary for polarization spectroscopy, lock-in detection significantly increases the signal to noise ratio of the measurement, and reduces the necessary probe beam power.  The lock-in amplifier was integrated into the MOGLabs DLC001 diode laser controller and operated at a fixed frequency of $250$ kHz.  We employ on/off square-wave amplitude modulation of the RF power driving the AOM with duty cycle adjusted to $50\%$ to maximize the lock-in signal. Lower-cost alternatives to our setup could incorporate mechanical beam choppers at the cost of lower modulation frequencies.

The AOM also had the effect of shifting the frequency of the pump beam by the RF driving frequency ($70$ MHz). The spectroscopic features therefore appear at the average frequency of the pump and probe beams, effectively producing a shift of $35$ MHz in our case. Double passing of the beam through the AOM and choosing the $+1$ and $-1$ diffraction orders would result in a net-zero frequency shift of the pump beam and allow the spectroscopic features to be centered at bare the transition frequencies.  

\section{Spectroscopic Measurements}\label{sec:spect_meas}

\begin{figure}[htbp]
\centering\includegraphics[]{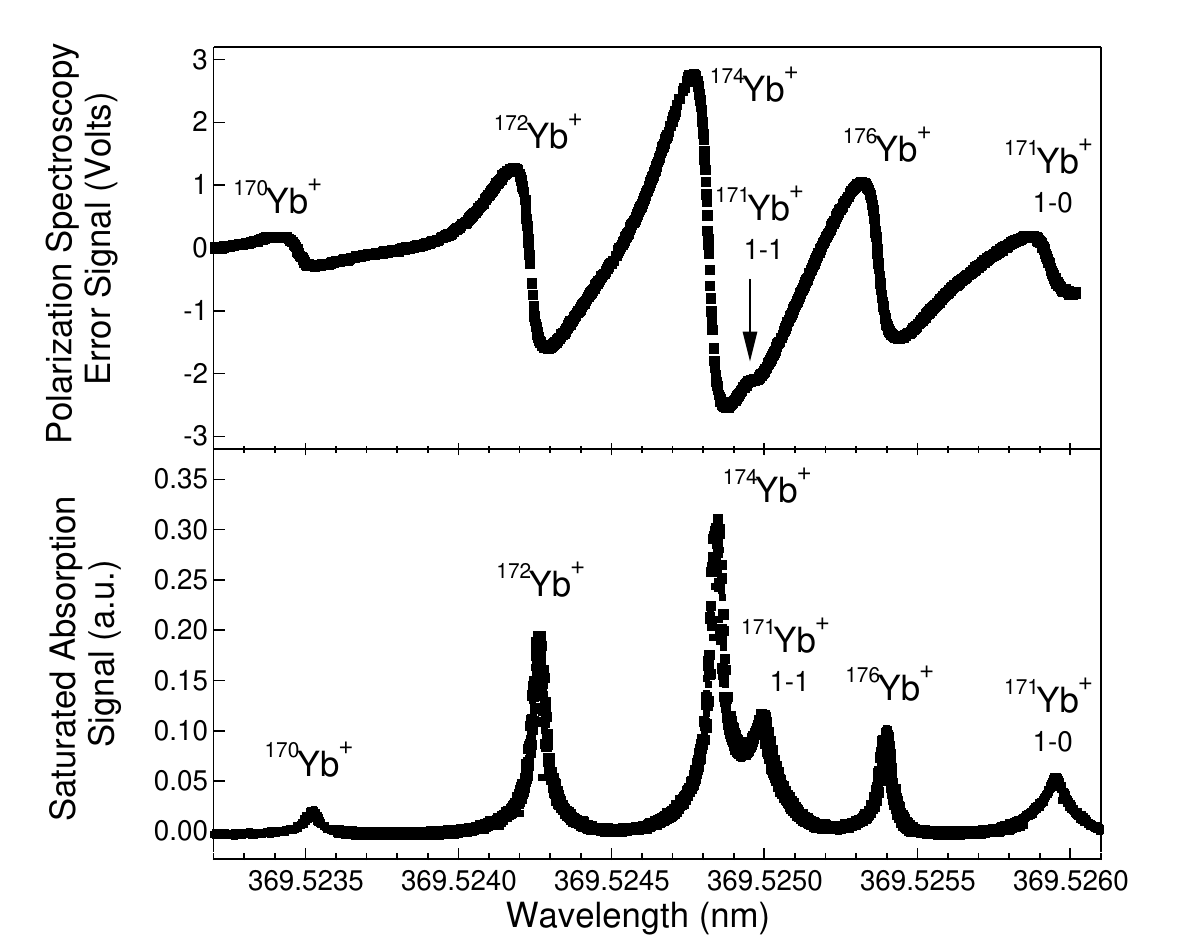}
\caption{The error signal generated by polarization spectroscopy (top) compared with the saturated absorption spectroscopy signal (bottom). Transition assignments made from \cite{nguyen2010ett,olmschenk2009qtb}. Isotopes with even atomic number each have a single feature while the 171 isotope shows features for both the $F = 1\rightarrow F' = 0$ and $F = 1\rightarrow F' = 1$ transitions. Transitions for the 173 isotope and the 171 $F = 0\rightarrow F' = 1$ transition lie outside the wavelength window of this measurement.  A systematic offset of approximately 50 MHz between the centers of the saturated absorption peaks and the zero-crossings of the polarization spectroscopy measurements are due to small wavemeter drifts between the measurements and the use of a double-pass AOM configuration in the saturated absorption spectroscopy, removing the 35 MHz frequency offset.}
\label{fig:pol_spec_sat_abs}
\end{figure}

Spectroscopic measurements were made by sweeping the ECDL wavelength over the range of the $^2S_{1/2}\rightarrow^2P_{1/2}$ transition of singly ionized Ytterbium ions of different isotopes as documented in previous work \cite{nguyen2010ett}.  The signals obtained for both polarization spectroscopy and saturated absorption spectroscopy conducted as per \cite{demtroder2008ls, pearman2002pso} are shown in Fig. \ref{fig:pol_spec_sat_abs}. Data were recorded by feeding the demodulated analog error monitor signal from the MOGLabs controller to a National Instruments data acquisition system, and simultaneously recording the laser wavelength from a HighFinesse WSU-10 wavemeter.  

In this wavelength range a number of features associated with $S-P$ transitions can be seen and associated with the different isotopes of Ytterbium following Nguyen \textit{et al} \cite{nguyen2010ett}.  The polarization spectroscopy measurements show the predicted dispersive lineshape that is ideal for laser locking applications with zero DC offset away from atomic transitions.  The saturated absorption measurements show features that correspond well to the polarization spectroscopy measurement in terms of both wavelength and relative size of each isotope's signal, with the signal size scaling proportionately with the natural abundance of the isotopes.  The transition for the $^{171}$Yb$^+$ ion, relevant to quantum information experiments, appears at the long-wavelength end of the spectrum.  

Typical peak-to-trough transitions in the polarization spectroscopy signal were measured to be $80-100$ MHz, much smaller than the Doppler-broadened transitions (several GHz). We observe that the presence of magnetic fields induced by proximal permanent magnets distorts the lineshape of the polarization spectroscopy signal but does not appear to change the zero-crossing location.

\section{Laser Frequency Stabilization}\label{sec:laser_lock}
The polarization spectroscopy measurements above provide a good frequency discrimination signal for laser frequency stabilization at a wavelength corresponding to the $^2S_{1/2}\rightarrow^2P_{1/2}$ transition in each of the isotopes of the Ytterbium ion. The dispersive type lineshapes are anti-symmetric about the transition wavelength and have a steep zero crossing at resonance. 

\begin{figure}[htbp]
\centering\includegraphics[]{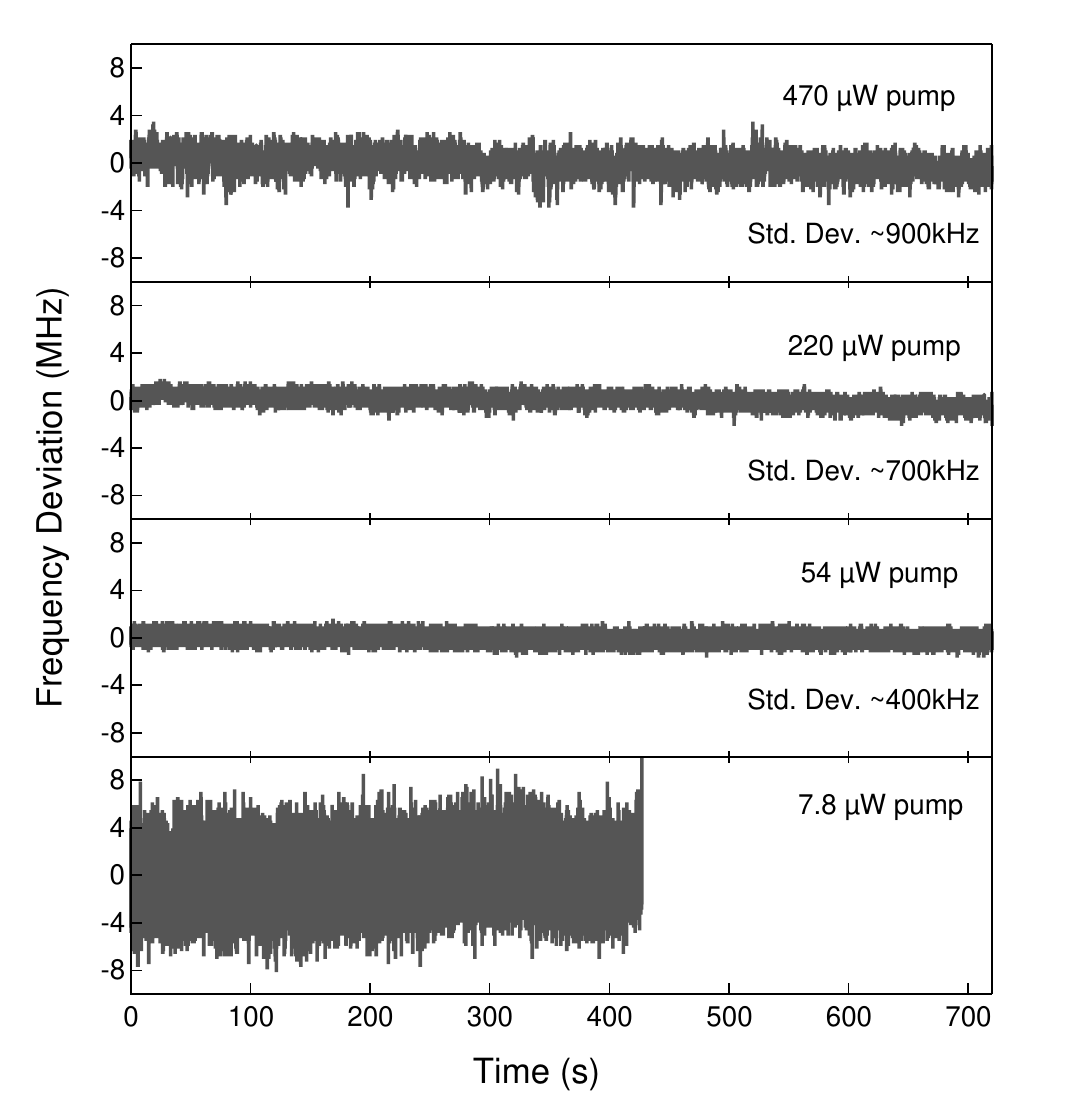}
\caption{Plot of wavelength stability measurements with varying power in the pump beam. The best frequency stability achieved was with 54 $\mu$W of pump power, where the standard deviation of the frequency was $\sim$400 kHz. Higher pump powers gave slightly worse stabiliy results, which could be due to photoemission from the pump beam impinging on the cathode material increasing with higher power and causing instability in the discharge, or from imperfect adjustment of the feedback servo with larger signals (the feedback gain was adjusted for best stability at each power level, however individual control of the PID parameters was not available). At the lowest pump power level of 7.8 $\mu$W the signal to noise ratio is too small and the ECDL drops from the lock during the test run. The frequency measuremetns were made with a HighFinesse WSU-10 wavelength meter set to the minimum 1ms integration time.}
\label{fig:stability_meas1}
\end{figure}

We are, in particular, interested in locking the laser to the $F = 1\rightarrow F' = 0$ hyperfine transition of the $^{171}$Yb$^+$ ion for laser cooling and state detection in ion-trap experiments.  Typical transition signals had a capture range of $\sim$100 MHz, much larger than typical frequency deviations in our laser due to, e.g. pressure changes or temperature drifts in the laboratory.  The slope of the dispersive signal can reach up to $0.01$ V/MHz (for $^{171}$Yb$^+$) using 510 $\mu$W in our probe beam.  The value of this parameter employed in this measurement was reduced from its maximum for optimal performance of our locking electronics.

Feedback was provided to the laser's piezo stack using the built-in PID circuitry in the MOGLabs diode laser controller. Independent gain controls are available to provide feedback to both the piezo ($\sim$1 kHz bandwidth) and the diode current ($\sim$40 kHz bandwidth) on different timescales using a single error signal.    Locking setpoints were adjustable using an analog offset voltage control incorporated into the laser diode controller.  We employed only slow feedback on the piezo stack in order to compensate for drifts in the free-running laser's output frequency, typically tens of MHz per minute.

The frequency stability of the locked laser was investigated by recording the laser frequency as a function of time using a wavelength meter.  The integration time on the wavelength meter was set to its minimum value of 1ms and the laser frequency stability was investigated by recording the stabilized laser frequency as a function of time for a variety of pump beam powers.

Plots of the frequency recordings are shown in Fig. \ref{fig:stability_meas1}. In all measurements the feedback stabilization removed the frequency drifts generally encountered with the laser in free-running mode.  The maximum deviations remained within a window of $\leq 5$ MHz peak-to-peak, as appropriate for efficient laser cooling on the 19 MHz $S-P$ transition in $^{171}$Yb$^+$.  

Interestingly, the measurement with the highest pump power of 470 $\mu$W does not display the best frequency stability, instead the best stability was observed with a pump power of 54 $\mu$W, where the standard deviation of the frequency over a $\sim$12 minute run was $\sim$400 kHz. As the pump power was reduced further and the signal-to-noise ratio of the differential measurement decreased, the locking stability eventually became degraded. We are not sure if the slightly reduced locking performance at the highest pump power was due to e.g. photoemission and instability in the discharge \cite{kumar1995poe,zhechev2005ict} or artifacts of the locking circuitry requiring manual optimization of the feedback gain for different error signal amplitudes.  Nonetheless, it is quite encouraging that such a stable lock could be achieved with only $\sim$50 $\mu$W of pump power.

\section{Discussion and Conclusion}

Experiments in atomic physics require lasers to be stabilized to within a few MHz of the transition frequencies for laser cooling, state preparation and detection; this usually requires some form of closed loop feedback based on a stable frequency reference. Broadly these frequency references can be placed into two categories, those based on an absolute frequency reference, such as atomic and molecular transitions and those based on calibrated frequency measurements, such as reference cavities or interferometers. Wavelength meters based on calibrated interferometers can have specifications such as absolute frequency accuracy of $\pm10$ MHz and precision of 0.1 MHz. Due to their ease of use and minimal optical power requirements of $\sim50$ $\mu$W they are attractive devices for laser frequency stabilization in their own right, particularly in the UV where relatively few absolute references exist. In practical laboratory setups however, the lock bandwidth of built-in PID regulators in these devices is often quite low (a few Hz or slower) and they require regular recalibration at the wavelength of interest in order to maintain the required frequency accuracy. 

On the other hand, schemes based on spectroscopy of atomic and molecular transitions may be insensitive to drifts in temperature and pressure or ageing of the apparatus. They also generally provide a significantly higher locking bandwidth, typically several 10s of kHz. Together these characteristics provide an effective means to damp external perturbations to the system. 

We have reported a simple and effective laser stabilization technique for experiments using trapped Yb$^+$ ions. Our approach used polarization spectroscopy on an absolute frequency standard in order to produce a dispersive lineshape compatible with laser frequency stabilization, ultimately reducing frequency deviations in a regulated laser to approximately the instantaneous linewidth of the laser. The optical setup employed was quite simple and did not require, e.g. multiple electro-optic modulators as in modulation transfer spectroscopy, or complex RF generators as in frequency-modulated saturated absorption. In our setup the use of an AOM could easily be replaced with an optical chopper, as on-off keying of the pump beam is the only requirement for lock-in detection.  

We have also generated Doppler-free dispersive lineshapes using frequency-modulated saturated absorption spectroscopy in a double-probe configuration \cite{king1999qse}.  However, this approach required significantly more challenging optical alignment and had more demanding technical requirements.  The two probes allowed common-mode laser amplitude noise rejection but required spatial separation within the $\sim2$ mm hollow-cathode aperture.  This in turn mandated beam diameters of approximately $200$ $\mu$m which greatly complicated probe overlap with only one beam. In addition, the size of the dispersive signal in this approach scaled with the modulation depth of the frequency dither on the pump.  Strong dispersive signals were generated with modulation depth of $\sim10$ MHz, meaning that a special broadband FM source allowing modulation up to $\sim15\%$ of the RF carrier on the AOM was required.  Qualitatively, we achieved more stable, lower-noise dispersive lineshapes using polarization spectroscopy.

Our approach is compatible with a wide variety of optical sources and frequency standards.  Hamamatsu provides hollow cathode lamps with cathode materials yielding dense absorption features across the UV and into the visible, from 193 nm to 422 nm.  This includes sources corresponding to other atomic species of interest for ion-trapping experiments including Barium, Beryllium, Calcium and Magnesium. As laser diode technology improves and direct sources in the UV become more readily available we believe these techniques will grow in utility for laser characterization and stabilization in ion trapping and atomic physics experiments.

\section*{Acknowledgements} 

The authors thank D. Kielpinski for advice on choice and operation of the HCL and S.D. Gensemer for technical assistance.  This work partially supported by the US Army Research Office under Contract Number  W911NF-11-1-0068, the Australian Research Council Centre of Excellence for Engineered Quantum Systems CE110001013, the Office of the Director of National Intelligence (ODNI), Intelligence Advanced Research Projects Activity (IARPA), through the Army Research Office, and the Lockheed Martin Corporation. All statements of fact, opinion or conclusions contained herein are those of the authors and should not be construed as representing the official views or policies of IARPA, the ODNI, or the U.S. Government.


\begin{thebibliography}{99}

\bibitem{metcalf1999lca} H. J. Metcalf and P. Van der Straten, \textit{Laser Cooling and Trapping.} (Springer, 1999).

\bibitem{arie1992afs} A. Arie, S. Schiller, E. K. Gustafson, and R. L. Byer, ``Absolute frequency stabilization of diode-laser-pumped Nd:YAG lasers to hyperfine transitions in molecular iodine,'' Opt. Lett. \textbf{17}, 1204-1206 (1992).

\bibitem{wallard1972fso} A. J. Wallard, ``Frequency stabilization of helium-neon laser by saturated absorption in iodine vapour,'' J. Phys. E: Scient. Instr. \textbf{5}, 926–930 (1972).

\bibitem{corwin1998fsd} K. L. Corwin, Z. T. Lu, C. F. Hand, R. J. Epstein, and C. E. Wieman ``Frequency-stabilized diode laser with the Zeeman shift in an atomic vapor,'' Appl. Opt. \textbf{37}, 3295-3298 (1998). 

\bibitem{ratnapala2004lfl} A. Ratnapala, C. J. Vale, A. G. White, M. D. Harvey, N. R. Heckenberg, and H. Rubinsztein-Dunlop, ``Laser frequency locking by direct measurement of detuning,'' Opt. Lett. \textbf{29}(23), 2704–2076 (2004).

\bibitem{hansch1980lfs} T. W. H\"{a}nsch and B. Couillaud, ``Laser frequency stabilization by polarization spectroscopy of a reflecting reference cavity,'' Opt. Commun. \textbf{35}(3), 441 (1980).

\bibitem{drever1983lpa} R. W. Drever, J. L. Hall, F. V. Kowalski, J. Hough, G. M. Ford, A. J. Munley, and H. Ward, ``Laser phase and frequency stabilization using an optical resonator,'' Appl. Phys. B \textbf{31}(2), 97–105 (1983). 

\bibitem{kielpinski2006lco} D. Kielpinski, M. Cetina, J. A. Cox, and F. X. Kärtner, ``Laser cooling of trapped ytterbium ions with an ultraviolet diode laser,'' Opt. Lett. \textbf{31}, 757-759 (2006).

\bibitem{nguyen2010ett} A. -T. Nguyen, L. -B. Wang, M. M. Schauer, and J. R. Torgerson, ``Extended Temperature Tuning of An Ultraviolet Diode Laser for Trapping And Cooling Single Yb+ ions,'' Rev. Sci. Instrum., \textbf{813}, 053110 (2010). 

\bibitem{wunderlich2003qma} C. Wunderlich and C. Balzer, ``Quantum measurements and new concepts for experiments with trapped ions,'' Adv. At. Mol. Opt. Phys. \textbf{49}, 293–376 (2003).

\bibitem{gill1995mot} P. Gill, H. A. Klein, A. P. Levick, M. Roberts, W. R. C. Rowley, and P. Taylor, ``Measurement of the $^{2} S_ {1/2}-^{2} D_ {5/2}$ 411-nm interval in laser-cooled trapped $^{172}$Yb$^{+}$ ions,'' Phys. Rev. A \textbf{52}, (1995).

\bibitem{sugiyama1995lco} K. Sugiyama and J. Yoda, ``Laser cooling of a natural isotope mixture of Yb$^+$ stored in an RF trap,''  IEEE Trans. Instrum. Meas. \textbf{44}, 140-143 (1995).

\bibitem{arie1993isa} A. Arie, M. L. Bortz, M. M. Fejer, and R. L. Byer, ``Iodine spectroscopy and absolute frequency stabilization with the second harmonic of the 1319-nm Nd:YAG laser,'' Opt. Lett. \textbf{18}, 1757–1759, (1993).

\bibitem{gerstenkorn1985dot}  S. Gerstenkorn and P. Luc, ``Description of the absorption spectrum of iodine recorded by means of Fourier transform spectroscopy: the $B–X$ system,'' Journal de Physique \textbf{46}, 867–881 (1985).

\bibitem{demtroder2008ls} W. Demtr\"{o}der, \textit{Laser Spectroscopy: Vol. 2: Experimental Techniques.} (Springer, 2008).

\bibitem{wieman1976dfl} C. Wieman and T. W. H\"{a}nsch, ``Doppler-free laser polarization spectroscopy,'' Phys. Rev. Lett. \textbf{36}, 1170-1173 (1976).

\bibitem{streed2008fso} E. W. Streed, T. J. Weinhold, and D. Kielpinski, ``Frequency stabilization of an ultraviolet laser to ions in a discharge,'' Appl. Phys. Lett. \textbf{93}, 071103-071103 (2008).

\bibitem{petrasiunas2012oso} M. J. Petrasiunas, E. W. Streed, T. J. Weinhold, B. G. Norton, D. Kielpinski, ``Optogalvanic spectroscopy of metastable states in Yb$^+$,'' Appl. Phys. B \textbf{107}, 1053–1059 (2012).

\bibitem{wang2009fso} W. Wang, J. Ye, M. Zhou and X. Xu, ``Frequency stabilization of a 399nm laser by modulation transfer spectroscopy in an ytterbium hollow cathode lamp,'' in \textit{Conference on Lasers and Electro-Optics/Pacific Rim} 2009, (Optical Society of America, 2009), paper TuB1.

\bibitem{pearman2002pso} C. P. Pearman, C. S. Adams, S. G. Cox, P. F. Griffin, D. A. Smith, and I. G. Hughes, ``Polarization spectroscopy of a closed atomic transition: applications to laser frequency locking,'' J. Phys. B \textbf{35}(24), 5141-5151 (2002).

\bibitem{ricci1995acg} L. Ricci, M. Weidem\"{u}ller, T. Esslinger, A. Hemmerich, C. Zimmermann, V. Vuletic, W. K\"{o}nig, and T. W. H\"{a}nsch, ``A compact grating-stabilized diode laser system for atomic physics,'' Opt. Commun. \textbf{117}, 541-549 (1995).

\bibitem{kumar1995poe} P. R. Sasi Kumar, V. P. N. Nampoori, C. P. G. Vallabhan, ``Photoemission optogalvanic effect near the instability region of a hollow cathode discharge,'' Opt. Commun. \textbf{118}, 525-528 (1995).

\bibitem{zhechev2005ict} D. Zhechev, N. Bundaleska, and J. T. Costello. ``Instrumental contributions to the time-resolved optogalvanic signal in a hollow cathode discharge,'' J. Phys. D: Appl. Phys. \textbf{38}, 2237–2243 (2005).

\bibitem{king1999qse} B. E. King, ``Quantum state engineering and information processing with trapped ions'' Ph. D. thesis, Department of Physics, University of Colorado, Boulder, (1999).

\bibitem{olmschenk2009qtb} S. Olmschenk, ``Quantum Teleportation Between Distant Matter Qubits,'' Ph. D. thesis, University of Maryland, (2009).
 
\end{thebibliography}
\end{document}